\documentclass{PoS}
\usepackage{graphicx}
\usepackage{epsfig}
\PoS{PoS(LAT2005)210}

\title{Radiative Transitions in Charmonium }

\ShortTitle{Radiative Transitions in Charmonium}

\author{Jozef Dudek, Robert Edwards, \speaker{David Richards}\\
        Jefferson Laboratory MS 12H2, 12000 Jefferson Avenue, Newport News, VA 23606, USA\\
E-mail: \email{dudek@jlab.org, edwards@jlab.org, dgr@jlab.org}}

\abstract{The form factors for the radiative transitions between
  charmonium mesons are investigated.  We employ an anisotropic
  lattice using a Wilson gauge action, and domain-wall fermion
  action.  We extrapolate the form factors to $Q^2 = 0$, corresponding
  to a real photon, using quark-model-inspired functions.  Finally,
  comparison is made with photocouplings extracted from the measured
  radiative widths, where known.  Our preliminary results find
  photocouplings commensurate with these experimentally extracted values.}

\FullConference{XXIIIrd International Symposium on Lattice Field Theory\

 25-30 July 2005\

 Trinity College, Dublin, Ireland}

\begin{document}

\section{Introduction}

The GlueX program at Jefferson Laboratory will produce hybrid mesons
through photoproduction, in contrast to the pion production mechanism
generally employed in other searches. This proposal is supported by
flux-tube model calculations, which show no suppression for
conventional-hybrid photocouplings compared with those for
conventional-conventional transitions\cite{Close:2003fz,Close:2003ae,PDBook}:
\begin{eqnarray}
\Gamma(\pi_{1H}^+ \to a_2^+ \gamma) & \sim &{\cal O}(100)~\mathrm{keV},\\
\Gamma(b_{JH}^+ \to \rho^+ \gamma) & \sim &{\cal O}(1000)~\mathrm{keV},\\
\Gamma(b_1^+ \to \pi^+ \gamma) & = & 230 \pm 60~\mathrm{keV}.
\end{eqnarray}

As a first examination of the validity of these calculations, we have
embarked on a program to examine transition form factors in the
charmonium sector.  Here we expect both a flux-tube model and lattice
calculations to provide reasonable descriptions of the physics.  In
this contribution, we focus on transitions between conventional
charmonium mesons as a precursor to future studies of hybrid mesons.
Many charmonium states below the $D \bar{D}$ threshold have measured
radiative transtions into ligher charmonium states, and these
transtions are reasonably well described in quark potential
models\cite{Eichten:2002qv}.

\section{Formalism}
We write the transition matrix element between an initial state $i$
and a final state $f$ as
\[
\langle f(\vec{p}_f, \lambda_f) \mid V_{\mu}(0) \mid
i(\vec{p}_i,\lambda_i) \rangle
\]
where $\vec{p}_i$ and $\vec{p}_f$ are the initial and final momenta
  respectively, $\lambda_i$ and $\lambda_f$ label helicities, and
  $V_{\mu}$ is the electromagnetic vector current.  For comparison
  both with experiment and with quark potential models, we express the
  transition form factors in a multipole expansion.  For the
  transitions we consider in this paper, the expansion is as follows:

\subsection{$\eta_c(0^-)$ form factor}
Whilst charge conjugation requires that this vanish, we can
nevertheless examine the ``form factor'' by coupling the vector
current only to the quark.  This is the paradigm process, analogous to
the pion form factor; continuum current conservation implies there is
only a single form factor:
\begin{equation}
\langle \eta_c (\vec{p}_f ) \mid V^{\mu}(0) 
\mid \eta_c ( \vec{p}_i ) \rangle = f(Q^2) \left[ p_f^{\mu} + p_i^{\mu}  
\right],
\end{equation}
where $-Q^2 = q^2 = (E_f - E_i)^2 - \vec{q}^2$, with $\vec{q} =
\vec{p}_f - \vec{p}_i$ the three-momentum carried by the
photon.

\subsection{$J/\psi (1^-) \longrightarrow \gamma \eta_c(0^-)$ transition}
This is an $M_1$-transition, and likewise described by only a single form
factor
\begin{equation}
\langle \eta_c( \vec{p}_{\rm PS}) \mid V^{\mu}(0) \mid 
\psi( \vec{p}_{\rm V}, \lambda)
\rangle = \frac{ 2 V(Q^2)}{m_{\eta_c} + m_\psi}
\epsilon^{\mu\alpha\beta\gamma} p_{\rm PS}^{\alpha} p_{\rm V}^{\beta}
\epsilon^\gamma(\vec{p}_{\rm V}, \lambda),
\end{equation}
where $\lambda$ is the helicity label of the $J/\psi$.

\subsection{$\chi_{c0}(0^+) \longrightarrow \gamma J/\psi(1^-)$ transition}
This transition involves an electric dipole $E_1(Q^2)$ coupling to the
transverse component of the photon, and a further form factor
$C_1(Q^2)$ coupling to the longitudinal component:
\begin{eqnarray}
\lefteqn{\langle \chi_{c0} (\vec{p}_{\rm S}) | V^\mu(0) | J/\psi
(\vec{p}_{\rm V}, \lambda) \rangle = } \nonumber \\ & &
\Omega^{-1}(Q^2) \Bigg( E_1(q^2) \left[ \Omega(Q^2)
\epsilon^\mu(\vec{p}_{\rm V}, \lambda) - \epsilon(\vec{p}_{\rm V},
\lambda).p_{\rm S} ( p_{\rm V}^\mu p_{\rm V}.p_{\rm S} - m_{\rm V}^2
p_{\rm S}^\mu \big) \right] + \nonumber\\ & & \left.
\frac{C_1(Q^2)}{\sqrt{q^2}} m_V \epsilon(\vec{p}_{\rm V}, \lambda).p_{\rm S}
\left[ p_{\rm V}.p_{\rm S} (p_{\rm V} +p_{\rm S})^\mu - m_{\rm S}^2
p_{\rm V}^\mu - m_{\rm V}^2 p_{\rm S}^\mu \right] \right)
\end{eqnarray}
For real photons, only $E_1$ is of interest, but in
our calculation we will extract both form factors.

\section{Computational details}
The computations are performed in the quenched approximation to QCD on
300 configurations of a $12^3 \times 48$ lattice.  We employ an
anisotropic Wilson gauge action~\cite{Klassen:1998ua}, with a
renormalized anisotropy $\xi \equiv a_s/a_t = 3$.  The temporal lattice spacing
obtained from the static quark-antiquark potential is $a_t^{-1} =
6.05 (1)~{\rm GeV}$.

The quark propagators are computed using an anisotropic domain-wall
fermion (DWF) action, with a domain-wall height $M = 1.7$ and $L_5 =
16$. We employ standard Gaussian smearing for the interpolating
fields, using sources of various widths.  For this exploratory study,
we do not attempt a precise tuning of the charm quark mass to yield
the physical $m_{\eta_c}$.  The charmonium spectrum obtained on our
lattices is listed in Table~\ref{tab:spectrum}; in general, the whole
spectrum is too light, reflecting the imprecision in our tuning.
\begin{table}
\begin{tabular}{r|lllll}
& $\eta_c$ & $J/\psi$ & $\chi_{c0}$ & $\chi_{c1}$ & $h_c$ \\ \hline
Lattice (MeV) & $2819(7)$ & $2917(7)$ & $3288(15)$ & $3401(29)$ &
$3351(19)$\\
PDG (MeV) & $2980(1)$ & $3097$ & $3415$ & $3511$ & $3526$\\
\end{tabular}
\caption{Determination of the spectrum, together with
the corresponding PDG
  values\protect\cite{PDBook}.}\label{tab:spectrum}
\end{table}

The three-point functions are computed using the usual
sequential-source method, following the strategy outlined in
ref.~\cite{Bonnet:2004fr}.  Sequential-source propagators are computed
for both a final scalar $\chi_{c0}$ and a pseudoscalar $\eta_c$, placed
at the mid-point of the lattice $t_f = 24$, for both $\vec{p}_f =
(0,0,0)$ and $\vec{p}_f = (1,0,0)$ where the momentum is expressed in
the appropriate lattice units.  The form factors are computed using
the local, non-conserved vector current.  The form factors are
obtained from the three-point functions with the remaining amplitudes
and masses extracted from fits to the two-point functions.  The
matching factor to the continuum, $Z_V$, is determined by imposing
charge conservation on the pseudoscalar form factor $f(Q^2 = 0) = 1$.
We find a discrepancy of around 11\% between the matching factor
obtained for a pion at rest, and that for a pion at the lowest
non-zero value of the lattice momentum $\vec{p} = (1,0,0)$, representing
an important systematic uncertainty on our calculation.

\section{Results}
\begin{figure}
\begin{center}
\psfig{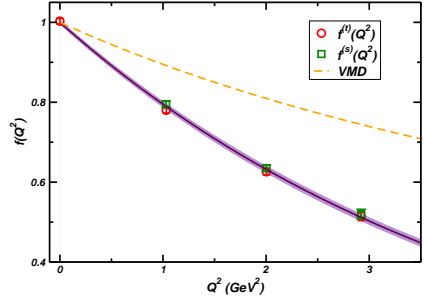}
\end{center}
\caption{The green and red data points show the lattice determination
  of the $\eta_c$ ``form factor'' obtained using the spatial and
  temporal components of the electromagnetic current respectively.
  The band corresponds to a quark-model-inspired fit to the data,
  while the dashed line is the VMD expectation, as described in the
  text.}
\label{fig:eta_ff}
\end{figure}

The $\eta_c$ ``form factor'' is shown in figure~\ref{fig:eta_ff}, for
the case $\vec{p}_f = (0,0,0)$.  In order to describe the
$Q^2$-dependence of the data, we appeal to a non-relativistic quark
model using harmonic-oscillator wave functions, characterized by a single
parameter $\beta$:
\begin{equation}
f(Q^2) = e^{ -Q^2/16 \beta^2}.\label{eq:gaussian}
\end{equation}
A single-parameter fit to the data yields $\beta = 0.522(5)~{\rm
  GeV}$, a value not uncharacteristic of that typically employed in
  quark-model charmonium wavefunctions; the fit is shown as the band
  in the figure.  Also shown as the dashed line is the VMD form, using
  the lattice value of the $J/\psi$ mass; VMD provides a
  notably poor description of the data, in contrast to the case of the
  pion\cite{Bonnet:2004fr}.

\begin{figure}
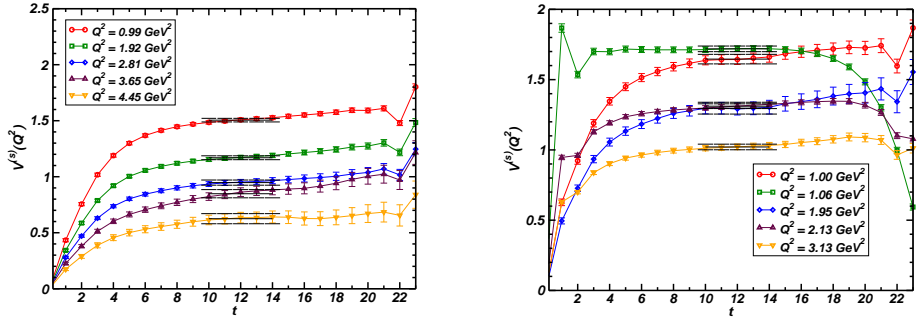

\begin{center}
\psfig{width=5.5cm,file=rho_pion_fit.out0_final.eps}\hspace{1cm}
\psfig{width=5.5cm,file=rho_pion_fit.out1_final.eps}
\end{center}
\caption{The left- and right-hand figures show plateaux in the
  determination of $V(Q^2)$ at several $Q^2$ for $\vec{p}_f = (0,0,0)$
  and $\vec{p}_f = (1,0,0)$, respectively.}
\label{fig:plateaux}
\end{figure}

Proceeding now to the transition form factors, we consider first the
transition $J/\psi \longrightarrow \gamma \eta_c$.  The quality of the
extraction of the form factor is illustrated in
Figure~\ref{fig:plateaux} where we show the plateaux at various $Q^2$
using the spatial components of the vector current.

\begin{figure}
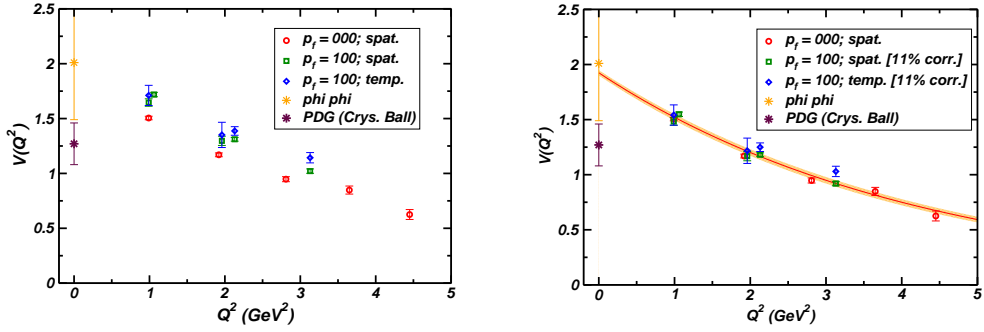

\begin{center}
\psfig{width=6.0cm,file=rho_pion_Qsq.eps}\hspace{1cm}\psfig{width=6.0cm,file=rho_pion_Qsq_with_fits.eps}
\end{center}
\caption{The left-hand plot shows the transition form factor $J/\psi
  \longrightarrow \gamma \eta_c$, for $p_f = (0,0,0)$ and $p_f =
  (1,0,0)$. The purple and yellow bursts are phenomenological values
  obtained from the PDG\protect\cite{PDBook}, and using the transition rate to
  $\phi \phi$ respectively.  The right-hand plot shows the same data
  with those corresponding to $\vec{p}_f = (1,0,0)$ scaled by the 11\%
  discrepancy in $Z_V$; the band is a quark-model-inspired fit to the
  lattice data, as described in the text.}
\label{fig:jpsi_eta}
\end{figure}

The $Q^2$ dependence of the transition form factor is shown as the
left-hand plot in Figure~\ref{fig:jpsi_eta}. The systematic difference
arising from the different determinations of $Z_V$ is manifest in the
discrepancy between the values obtained with $\vec{p}_f = (0,0,0)$ and
those with $\vec{p}_f = (1,0,0)$; the right-hand plot shows the data
with those for $\vec{p}_f = (1,0,0)$ rescaled by this 11\%, greatly
reducing this discrepancy. In order to extrapolate to $Q^2 =
0$, we once again employ a non-relativistic quark-model-inspired form
\begin{equation}
V(Q^2) = V(0) e^{-Q^2/16 \beta^2}.
\end{equation}
A combined fit to the $\vec{p}_f = (0,0,0)$ and rescaled $\vec{p}_f = (1,0,0)$
data yields $\beta = 0.52(1)~{\rm GeV}$ and $V(0) = 1.92(2)$.

The radiative width is related to the photocoupling $V(Q^2 = 0)$ by
\begin{equation}
\Gamma(J/\psi \to \eta_c \gamma) = \alpha \frac{|\vec{q}|^3}{(m_\psi+m_{\eta_c})^2} \frac{4}{3} |V(Q^2=0)|^2,
\end{equation}
whence we can obtain an ``experimental'' determination of the
photocoupling using the measured radiative width.  This is shown for
the PDG value as the purple burst\cite{PDBook}; an alternative value
obtained from the ratio of the product branching fraction of $\psi
\longrightarrow \gamma \eta_c \longrightarrow \gamma \phi \phi$ and
the $\eta_c \longrightarrow \phi \phi$ branching fraction is shown in
yellow.

\begin{figure}
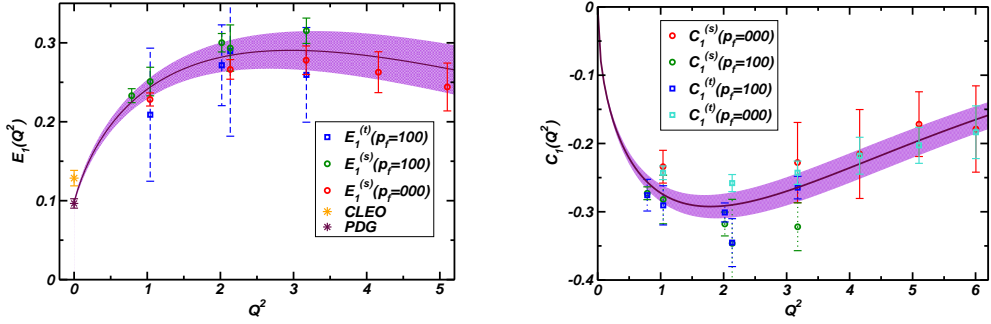

\begin{center}
\psfig{width=6.0cm,file=rho_a0_E1_Qsq.eps}\hspace{1cm}
\psfig{width=6.0cm,file=rho_a0_C1_Qsq.eps}
\end{center}
\caption{The left- and right-hand plots show the form factors
  $E_1(Q^2)$ and $C_1(Q^2)$ respectively for the radiative transition
  $\chi_{c0} \longrightarrow \gamma J/\psi$.  The bands correspond to
  independent, quark-model-inspired fits to the lattice data at
  $\vec{p}_f = (0,0,0)$ and $\vec{p}_f = (1,0,0)$, as described in the
  text.  The purple and yellow bursts are phenomenological values
  obtained from the PDG\protect\cite{PDBook}, and from a recent CLEO
  determination\protect\cite{Adam:2005uh}}
\label{fig:e1_ff}
\end{figure}

Finally, we turn to the $\chi_{c0} \longrightarrow \gamma J/\psi$
transtion.  The two form factors $E_1(Q^2)$ and $C_1(Q^2)$ are shown
as the left- and right-hand plots in Figure~\ref{fig:e1_ff}.  Using
a quark-model form for the extrapolation to $Q^2 = 0$, we have for the
electric-dipole form factor
\begin{equation}
E_1(Q^2) = c \mid \vec{q}(Q^2) \mid e^{-Q^2/16 \beta^2}
\end{equation}
where $\vec{q}$ is the three-momentum of the photon in the rest frame
of the decaying $\chi_{c0}$,
\begin{equation}
\mid \vec{q}(Q^2)\mid^2 =
\frac{(m_\psi^2 - m_\chi^2)^2 + 2 Q^2 (m_\psi^2 + m_\chi^2) + Q^4}{4 m_\chi^2},
\end{equation}
with an analogous form for the longitudinal multipole $C_1(Q^2)$.  The
extrapolations are shown as the bands on the figure.  Once again, we
can use the measurement of the radiative width to obtain an
``experimental'' value for the physical multipole $E_1(Q^2 = 0)$. This
is shown as the purple burst using the PDG width\cite{PDBook}, and as
the yellow burst using a recent CLEO determination\cite{Adam:2005uh}.
The corresponding values for the parameter $\beta$ are $0.60(3)~{\rm
GeV}$ and $0.47(1)~{\rm GeV}$ for the $E_1$ and $C_1$ multipoles
respectively.

In this work, we present a first study of radiative transitions
in charmonium.  While the results are preliminary, we 
find quark-model extrapolations yield values for the photocouplings
in reasonable proximity to experimental expectations, with broadly
consistent values for the wave-function parameter $\beta$.
Computations of further form factors are in progress.  A future
goal is to extend this study to the light-quark sector relevant
for JLab, and to transitions involving hybrid mesons.

\section*{Acknowledgements}
This work was supported by DOE contract DE-AC05-84ER40150 under which
the Southeastern Universities Research Association operates the
Thomas Jefferson National Accelerator Facility.  We are grateful for
discussions with Ted Barnes, Frank Close, George Fleming and Jim
Napolitano.

\bibliography{writeup}


\end{document}